\documentclass[10pt,conference]{IEEEtran}
\IEEEoverridecommandlockouts
\usepackage{cite}
\usepackage{csquotes}
\usepackage{amsmath,amssymb,amsfonts}
\usepackage{algorithmic}
\usepackage{graphicx}
\usepackage{textcomp}
\usepackage{xcolor}

\usepackage{booktabs,tabularx,makecell}
\usepackage{todonotes}
\usepackage[hidelinks]{hyperref}

\def\BibTeX{{\rm B\kern-.05em{\sc i\kern-.025em b}\kern-.08em
    T\kern-.1667em\lower.7ex\hbox{E}\kern-.125emX}}
    
\usepackage{tikz}

\begin{document}

\definecolor{colorA}{RGB}{114,141,196}
\definecolor{colorB}{RGB}{214,104,99}
\definecolor{colorC}{RGB}{109,179,147}
\definecolor{colorD}{RGB}{199,123,227}
\definecolor{colorE}{RGB}{255,193,116}
\definecolor{colorF}{RGB}{179,140,121}

\title{Teaching Computer Science Students to Communicate Scientific Findings More Effectively}

\author{
	\IEEEauthorblockN{Marvin Wyrich and Stefan Wagner}
    \IEEEauthorblockA{
    Institute of Software Engineering,
    University of Stuttgart\\
    Stuttgart, Germany\\
    \{firstname.lastname\}@iste.uni-stuttgart.de
    }
}

\maketitle

\begin{abstract}
Science communication forms the bridge between computer science researchers and their target audience. Researchers who can effectively draw attention to their research findings and communicate them comprehensibly not only help their target audience to actually learn something, but also benefit themselves from the increased visibility of their work and person.
However, the necessary skills for good science communication must also be taught, and this has so far been neglected in the field of software engineering education.

We therefore designed and implemented a science communication seminar for bachelor students of computer science curricula. Students take the position of a researcher who, shortly after publication, is faced with having to draw attention to the paper and effectively communicate the contents of the paper to one or more target audiences. Based on this scenario, each student develops a communication strategy for an already published software engineering research paper and tests the resulting ideas with the other seminar participants.

We explain our design decisions for the seminar, and combine our experiences with responses to a participant survey into lessons learned. With this experience report, we intend to motivate and enable other lecturers to offer a similar seminar at their university. Collectively, university lecturers can prepare the next generation of computer science researchers to not only be experts in their field, but also to communicate research findings more effectively.
\end{abstract}


\section{Introduction}
Impact.
If there is one term that comes up again and again when it comes to evaluating scientific research and its authors, it is \emph{impact}.
It does not seem to matter whether a software engineering researcher feels closest to mathematicians, social scientists, or engineers:\footnote{An allusion to Lionel C. Briand's ICSE 2022 keynote on \textit{the split minds of software engineering researchers}: https://tinyurl.com/splitMinds} all researchers want the results of their work to at least reach the intended target group.

It is all the more remarkable that the work of a software engineering researcher usually ends with the publication of a paper -- so to speak, at the moment when the greatest possible impact could be generated in the first place.
Although conference publications are usually accompanied by an obligatory presentation of the research paper and the papers themselves are often well written, in reality this is not sufficient to reach the potentially interested target audience:
Conference presentations are usually only accessible to a small part of the target group, for example, due to local or time restrictions.
Plus, we can all still improve a lot when it comes to how we design our presentations to convey knowledge more effectively and stay positively in the minds of our audiences.
The scientific papers themselves are also only accessible to a few people, either because they are published behind paywalls or simply because non-scientists have no routine in reading scientific literature.
On top of this comes that those who would have access must first become aware of the new research findings.

In breakout sessions at various seminars, we discussed with members of the software engineering research community what their understanding of science communication is.
All the answers fell into two categories: first, drawing attention to and disseminating scientific findings, and second, establishing oneself as a scientist.
The former aspect fits well with the definition of Burns et al.:
\enquote{Science communication (SciCom) is defined as the use of appropriate skills, media, activities, and dialogue to produce one or more of the following personal responses to science: Awareness, Enjoyment, Interest, Opinion-forming, and Understanding}~\cite{Burns:2003:SciComDefinition}.

Depending on the current demand for the research topic, the combination of expertise and good science communication can then result in great publicity for one's own person.
For example, during the Corona pandemic, virologists in some countries became talk show celebrities, whether intentionally or not.
Of course, this does not have to be the goal of every software engineering researcher.
Establishing oneself within one's own research community is also more often the case than becoming a person of public interest.

In any case, we argue, as do some of our peers~\cite{Brownell:2013:SciComTeachStudents}, that good science communication can and should be learned in science education.
And since both the researchers themselves and their target audiences would benefit from better science communication, we further argue that it should be taught as early as possible.
We therefore decided to offer a science communication seminar for bachelor students of computer science curricula.
Since both, us teachers and the students, were very satisfied with the result of the first seminar implementation, we would like to share our experiences and lessons learned in this paper.
We provide all information needed to successfully implement similar science communication seminars at other universities, enabling the next generation of computer science researchers to effectively bridge the gap between academia and their target audience.

\begin{figure*}[t]
    \centering
    \includegraphics[width=1\textwidth]{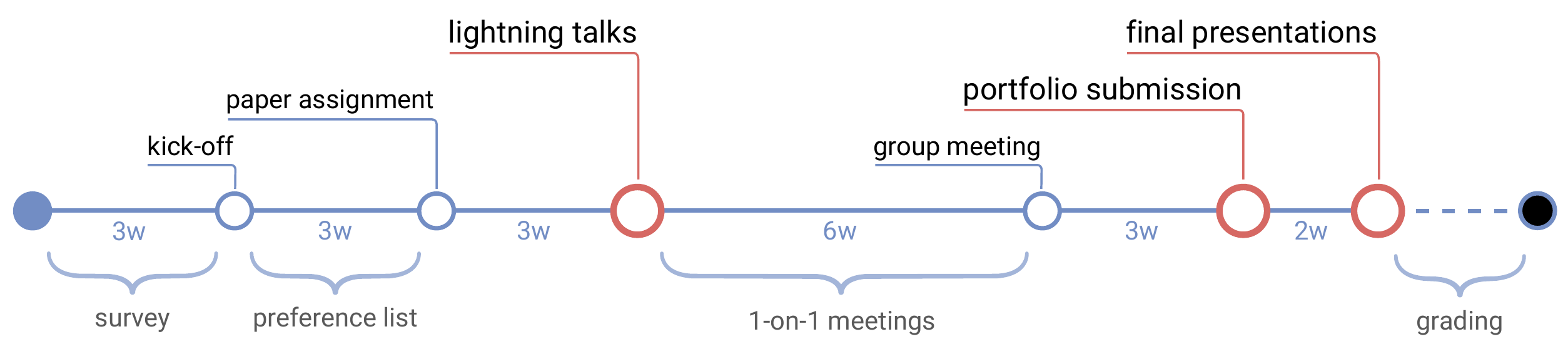}
    \caption{Timeline of the seminar from start of semester to completion of seminar}
    \label{fig:overall_structure}
\end{figure*}

\section{Course Design}

We have designed the course as a 6-month seminar for bachelor students from computer-science-related study programs (e.g., software engineering, media informatics, computer science).
Students in those programs must choose one such graded seminar worth 3 ECTS credits ($\approx 90$ study hours) from a list of seminars.

The idea of the seminar is simple: each student is assigned an already published scientific software engineering paper, which they will deal with intensively during the seminar.
In practical terms, this means that a portfolio of worked-out ideas for the communication of the paper contents must be handed in at the end of the seminar.
Students take the position of a researcher who, shortly after publication, is faced with having to draw attention to the paper and effectively communicate the contents of the paper to one or more target audiences.

\subsection{Overall Structure}

Fig.~\ref{fig:overall_structure} depicts the timeline from the beginning of the semester to the completion of the seminar at the end of the lecture period.
Circles on the timeline represent milestones, with the bigger red ones representing a milestone at which students had to deliver outcomes.

The seminar began with a welcome email from the course instructor to the seminar participants, referring to the course forum\footnote{We used ILIAS, an open-source learning management system that our students were already familiar with. Any comparable forum where materials can be shared and questions asked would work. \url{https://www.ilias.de/en/}} and the upcoming kick-off meeting.
Before the kick-off, students were asked to take a few minutes to write something in a Padlet\footnote{A digital pinboard where participants could add content in predefined structures (in this case, columns with predefined headings): \url{https://padlet.com/}} about their background and expectations for the seminar.
The results were then picked up at the kick-off meeting.

In the three weeks following the kick-off, each student submitted a preference list of papers that were up for selection and was then assigned a paper by the instructor.
Another three weeks later, each participant gave two-minute lightning talks on the contents of their paper.

From then on, the phase in which the students worked on their communication strategy portfolio began.
There were individual 1-on-1 meetings between instructor and student, as well as an informal group meeting after 6 weeks.
A total of 9 weeks after the lightning talks, the portfolios had to be handed in.
Part of the portfolio was a mandatory chapter for the design of the final presentation, allowing the students to use the two weeks after the portfolio submission deadline to refine and practice their already drafted presentation. 
The final presentations concluded the science communication seminar for the students.

In the following, we elaborate on details of individual milestones and course characteristics.

\subsection{Learning Objectives and Principles}
\label{sec:learning_objectives}

The course had two learning objectives:

\begin{enumerate}
    \item Students should be educated about the scientific process and working with scientific literature. Many of the participants have little experience in working with scientific literature, but will need these skills for their bachelor thesis at the latest. The introduction to the questions \enquote{what science says} and \enquote{how scientists do things}, i.e. science education, has overlapping goals with science communication, and can therefore be combined very well~\cite{Baram-Tsabari:2015:BridgingEduSciCom}.
    \item Students can reflect on the state of the art in communicating scientific software engineering findings and should be encouraged to try things that might be beneficial to science communication, even if they seem unconventional at first. 
\end{enumerate}

We emphasize that the course is not primarily about telling students what good science communication practices are, nor about turning each of the participants into a good speaker.
Instead, we intend to enable them to question the status quo and to reflect for themselves on how presentations and other means of communication should be designed to effectively communicate scientific knowledge.
We show the students possible boundaries through insights into scientific processes.
Within these boundaries, they have a lot of room to experiment according to their own strengths and weaknesses and receive feedback from the other seminar participants.

At the same time, we do not want them to be innovative just for being innovative.
We make it clear early on that we are evaluating the students' discussion of the potential impact of their ideas; not their level of innovation.
If they use established methods to effectively communicate knowledge, and can justify why these methods are potentially effective, they have done everything right.

We further emphasize that we are well aware that statistically a large part of the seminar participants will not end up pursuing an academic career, but that the course is nevertheless valuable for students.
The skills learned will at least help them in their final thesis, and the ability to handle scientific literature can make them better computer scientists overall.
In addition, being able to communicate work results in an understandable way is also rewarding in an industry job.

\subsection{Participants and Instructors}

Eight bachelor students registered for our seminar.
One student had to quit in the middle of the semester due to a change of study program, the remaining seven successfully completed the seminar.

After this first seminar run, we believe that the ideal group size for the seminar is between six and twelve participants.
By having a group size that is not too small, one learns about diverse personal preferences and styles of knowledge communication.
By having a group that is not too large, where students get to know each other over time, they are more likely to have the courage to venture out of their comfort zone and explore new presentation methods.

Our course had one instructor and one official examiner.
The instructor handled all communication with and supervision of the students, as well as moderated all group meetings (kick-off, lightning talks, intermediate group meeting, final presentations).
The examiner took part in the lightning talks and final presentations to participate in the feedback rounds.
At the end, the instructor and the examiner together decided on the seminar grades (see~\ref{sec:grading}).

The instructor should have an interest in science communication and be able to reflect on and justify how a certain type of knowledge communication can affect different target groups.
The authors themselves have such an interest, but would not call themselves experts in the field, nor do they think it is essential to be an expert to run the course.
It is more important that both instructor and examiner can admit other opinions and discuss with students on an equal footing.
Otherwise, there is a risk that students will not be encouraged to question the status quo in science communication established by the previous generation of researchers.
Our recommendation for the feedback rounds after a lightning talk or final presentation is that the instructor and the examiner should hold back at first and let the students ask their questions and provide feedback on the presentation.

\subsection{Kick-Off Meeting}

The kick-off meeting marked the official start of the seminar.
Students were told in advance that they only needed to fill out a Padlet in advance for this meeting, but otherwise no preparation was necessary.
The Padlet contained 7 columns with predetermined headings that prompted the student to write something in most columns.
We asked for expectations and concerns about the upcoming seminar, as well as examples of well-known people and channels that are great at communicating knowledge.
After the instructor briefly introduced himself, and explained the motivation behind the seminar, the instructor and students went over the students' Padlet responses together.

We complemented the examples of good knowledge communication channels with a brief collection of resources specifically designed to improve science communication in general, and some that already provide examples of what we believe to be good science communication channels in computer science.\footnote{See, e.g., \url{https://neverworkintheory.org} and \url{https://chuniversiteit.nl/papers}.}
This concluded the first half of the kick-off meeting, and all participants now had a rough idea of the seminar's intention.

In the second half of the meeting, we explained to the students what was expected of them and that we would guide them step by step through the seminar.
For the students, the most important thing at that moment was to understand that in the seminar, everyone has to deal with exactly one already published paper and that the selection of this paper was now the next step.

\subsection{Paper Assignment}

We asked members of the SE research community via Twitter for suggestions for papers for such a seminar and received a handful of good suggestions.
We initially selected 10 software engineering papers based on the following criteria:

\begin{itemize}
    \item The papers are neither authored by the instructor, the examiner, nor by other members of our research group to avoid bias.
    \item The papers are of comparable length and about the length of a typical conference full paper.
    \item The papers cover topics that require little prior knowledge to understand.
\end{itemize}

For each of the 10 papers, each student now had to choose one of three options: \enquote{Top 3} if the paper is a personal favorite, \enquote{Would be okay} if the student is interested in the paper, and \enquote{Rather not} if the paper does not interest the student at all.
The \enquote{Top 3} option could be assigned to a maximum of three papers, while the other two options could be selected as often as needed.

Table~\ref{tab:seminar_papers} lists the papers that were available for selection and shows the results of the preference allocation of eight students.
All but one paper landed in at least one student's personal top 3.
Furthermore, there were only three papers for which the \enquote{Rather Not} option was assigned by a majority.
This suggests that the students were mostly satisfied with the initial paper selection.
Since their votes were also favorably distributed, we were able to assign each student a paper that was among his or her favorites.
Of course, this can be different with other seminar participants and other papers.
Nevertheless, for reasons of motivation, we should always try to ensure that no student has to deal with a paper that he or she has marked as \enquote{Rather not}.

\begin{table*}[th]
\caption{List of the 10 papers offered for selection and the per paper aggregated preference results of eight seminar participants \\(\enquote{\textcolor{colorA}{Top 3}}, \enquote{\textcolor{colorE}{Would be okay}}, \enquote{\textcolor{colorB}{Rather not}})}
\centering
\label{tab:seminar_papers}
\begin{tabularx}{\textwidth}{l X l}
\toprule
Paper Title & & Preferences \\
\midrule
Debugging Hiring: What Went Right and What Went Wrong in the Technical Interview Process & \cite{Behroozi:2020:Hiring} & 
  \def\preftop{62}%
  \def\prefokay{25}%
  \def\prefnot{13}%
  \input{_prefBar}%
 \\
\enquote{How Was Your Weekend?} Software Development Teams Working From Home During COVID-19 & \cite{Miller:2021:Weekend} & 
  \def\preftop{62}%
  \def\prefokay{25}%
  \def\prefnot{13}%
  \input{_prefBar}%
\\
Hashing It Out: A Survey of Programmers' Cannabis Usage, Perception, and Motivation & \cite{Endres:2022:Hashing} & 
  \def\preftop{38}%
  \def\prefokay{49}%
  \def\prefnot{13}%
  \input{_prefBar}%
 \\
Reel Life vs. Real Life: How Software Developers Share Their Daily Life through Vlogs & \cite{Chattopadhyay:2021:ReelvsReal} & 
  \def\preftop{38}%
  \def\prefokay{37}%
  \def\prefnot{25}%
  \input{_prefBar}%
 \\
I Know What You Did Last Summer - An Investigation of How Developers Spend Their Time & \cite{Minelli:2015:IKnowWhat} & 
  \def\preftop{25}%
  \def\prefokay{62}%
  \def\prefnot{13}%
  \input{_prefBar}%
 \\
The Power of Bots: Characterizing and Understanding Bots in OSS Projects & \cite{Wessel:2018:Bots} & 
  \def\preftop{25}%
  \def\prefokay{25}%
  \def\prefnot{50}%
  \input{_prefBar}%
 \\
Objects count so count objects! & \cite{Tempero:2018:Objects} & 
  \def\preftop{25}%
  \def\prefokay{13}%
  \def\prefnot{62}%
  \input{_prefBar}%
 \\
Why don't software developers use static analysis tools to find bugs? & \cite{Johnson:2013:WhyStatic} & 
  \def\preftop{13}%
  \def\prefokay{62}%
  \def\prefnot{25}%
  \input{_prefBar}%
 \\
Program Comprehension and Code Complexity Metrics: An fMRI Study & \cite{Peitek:2021:fMRI} & 
  \def\preftop{13}%
  \def\prefokay{12}%
  \def\prefnot{75}%
  \input{_prefBar}%
 \\
The Product Backlog & \cite{Sedano:2019:Backlog} & 
  \def\preftop{0}%
  \def\prefokay{13}%
  \def\prefnot{87}%
  \input{_prefBar}%
\\
\bottomrule
\end{tabularx}
\end{table*}

\subsection{Lightning Talks}

After each student was assigned a paper, they had three weeks to prepare a lightning talk.
We instructed the students to understand their paper as best they could during this phase and to take notes on their first impressions of the paper, such as what they thought about the content and what they thought about how the paper authors presented their work.
They might come back to their first impressions later when they discuss the paper presentation of the original authors as part of their portfolio.

The goal of each lightning talk was to draw attention to the paper and explain what it was about at its core.
The students had to decide for themselves what they wanted to focus on.
This could be, e.g., the motivation behind the study, or its results, or a combination of different aspects.
The twist, however, was that a student had no more than 2 minutes for their lightning talk.
We deliberately chose the time limit to be short so that the contents of the paper could not be described in detail, thus provoking questions from the audience.
This was to help the presenter understand what was of interest to the audience about the topic, methodology, or other aspects of the paper.
Such cues will be useful later in developing the communication strategy.

Once the two-minute talk was over, the presenter could take a short rest.
The other seminar participants now had time to write down their questions and feedback on the presentation using an online tool.
This made it possible for the presenter to read through the entire feedback at a later time, without having to take notes on each question.
Some feedback and questions were then asked live by the seminar participants to the presenter before the next lightning talk started.

\subsection{Portfolio}

The focus of the seminar was for each student to create a portfolio.
A portfolio in this seminar was a collection of developed and discussed ideas for communicating the contents of the assigned paper.
It can be thought of as a report consisting of several chapters in which the ideas are described and discussed.
Therefore, the final deliverable was primarily a PDF with possibly additional artifacts such as short videos or poster designs.

For the portfolio, we specified mandatory parts:
\begin{itemize}
    \item An introduction that explains the motivation for the portfolio and provides an overview of the portfolio.
    \item A factual summary of the paper of at least one and no more than two pages.
    \item An illustration of the paper in a single figure.
    \item A summary of the paper, explicitly addressed to non-scientists and not longer than 200 words.
    \item A discussion of the paper. The focus here depends on the specific paper, but in general, it could be interesting to assess the presentation of the paper, i.e. how could the paper have been better prepared to be more accessible to readers, and e.g., to which target group the contents of the paper are most likely to be directed (software developers? team leaders? IT admins? scientists? \ldots).
    \item Design of structure and content of the final presentation.
\end{itemize}

Furthermore, each student should decide on one or more additional ideas.
We emphasized that we can discuss anything that could contribute to science communication, even if a suggestion seems unconventional at first.
Inspiration for ideas can be found, e.g., in the works of Illingworth~\cite{Illingworth:2017:DeliveringEffective} and Cooke et al.~\cite{Cooke:2017:Considerations}.
Their suggestions range from science festivals to book clubs, but also contain some general advice on effective knowledge communication.
In our first seminar implementation, students submitted short videos, Instagram Stories, blog posts, news articles, workshop concepts, and posters.

An important note that we have also expressed to students: There are entire degree programs where students spend several years learning how to develop a communications strategy.
We do not presume to act on a similar level; our seminar is not about developing the perfect video or the most polished workshop concept.
Rather, it is to outline and implement some ideas prototypically, and then discuss primarily their potential effectiveness and drawbacks in science communication (see section~\ref{sec:learning_objectives} on learning objectives).
We felt that this explicit information took some pressure off the computer science students and increased their joy of experimenting in unfamiliar territory.
We encouraged them, where possible, to implement (prototypically) the strategies they came up with.

\subsection{Intermediate Meeting(s)}

Each student was encouraged to approach the instructor in the first few weeks after the lightning talk and schedule an informal 1-on-1 meeting to discuss ideas and progress on the portfolio.
Most of our students took advantage of this offer.

In addition, an informal group meeting was held six weeks after the lightning talks.
To break the ice, everyone began by briefly telling the group about his or her current status in portfolio development.
Most of the time, questions or uncertainties arose that could be discussed in the group.
Then the instructor addressed a few points that came up repeatedly in the 1-on-1 meetings and could be interesting for all participants.
Finally, we discussed the procedure for the final presentations, which were scheduled five weeks after this group meeting.

\subsection{Final Presentations}

For the final presentations, all students, the instructor and the examiner came together for the last time.
Each student was granted 15 minutes on stage to present about their results.

In contrast to the lightning talk, the focus of the final presentation was no longer exclusively on the content of the paper, but also on the communication strategy that the student had developed.
Ideally, parts of the developed communication strategy could be used to present the contents of the paper.
For example, one student engaged the audience in a role-play right at the beginning, welcoming them as participants in a workshop for managers on improving job satisfaction in the home office.
The student continued in the role of the workshop leader for a while, until he explained that this was how the audience could imagine the entry to a workshop he had designed for his paper on software development teams working from home during COVID-19~\cite{Miller:2021:Weekend}.

Following each presentation, the other seminar participants asked questions and provided constructive feedback on the presentation and the developed communication strategy.

\subsection{Grading}
\label{sec:grading}

We designed our course according to the principle of constructive alignment~\cite{Biggs:1996:ConstructiveAlignment}, which means that we aligned our teaching activities as well as student assessment with the learning objectives.
As a quick reminder, the course had two learning objectives.
First, by the end of the seminar, students should be better informed about the scientific process and have gained practical experience in working with scientific literature.
Second, students should be able to reflect on how to communicate scientific knowledge effectively.

The assessment focused on the submitted portfolio and, in particular, on the discussion of the effectiveness of individual communication methods.
We consider the first learning objective to be fulfilled if the content of the assigned paper is correctly conveyed in the communication strategy and if certain developed ideas or criticisms of the paper authors' communication take into account the current constraints of the scientific process and paper format specifications.
We consider the second learning objective to be met when the student makes a comprehensible argument for the potential benefits and potential limitations of each portfolio component.

It is in the nature of all evaluations that there is an inherent subjectivity to them.
The particular difficulty with design work is that the amount of work invested by the student, which an examiner may wish to include in the assessment, is not always evident in the outcome.
This may be the case, first, if the creative process involves learning from discarding earlier ideas.
Second, it may be that the solution that is actually the simplest is also the most effective in the end.
Therefore, we evaluated the portfolios by assessing the students' reasoning about the communication concept they had developed and by assessing the quality of implementation, but without (consciously) speculating on the amount of work invested.

\section{Impressions of the Resulting Communication Strategies}

The seminar participants gave us their consent to present excerpts from their portfolios in anonymized form.
We will do so in this section, as the results have turned out nicely and concrete impressions complement the previous description of the seminar well.

\begin{figure*}[t]
    \centering
    \includegraphics[width=1\textwidth]{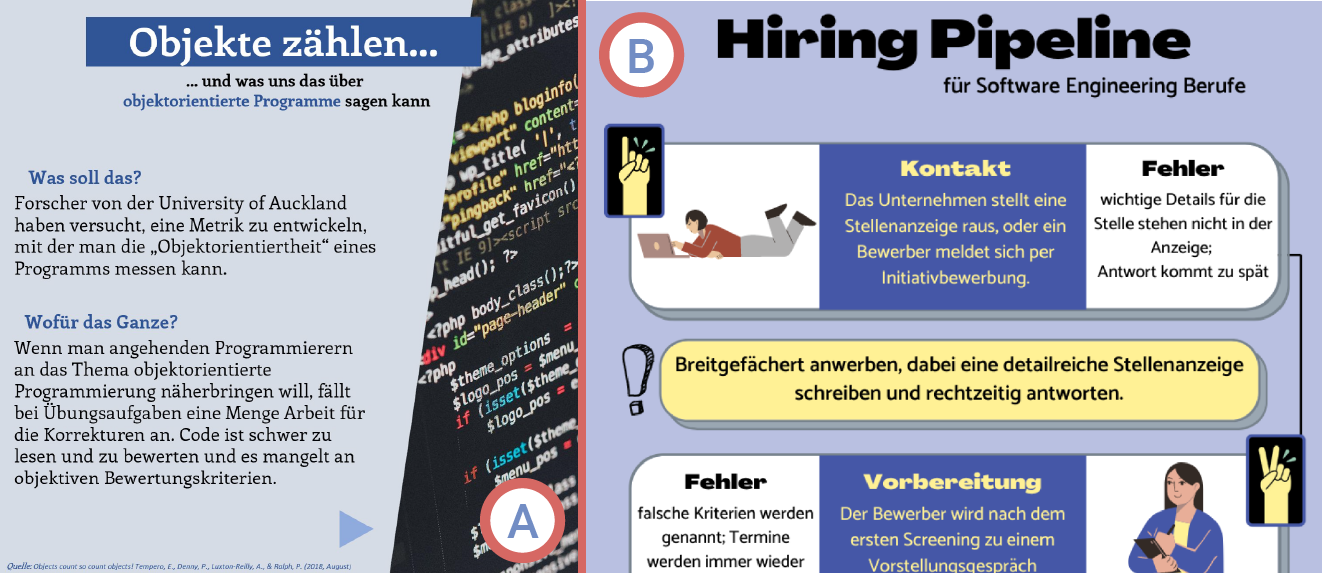}
    \caption{Impressions on communication artifacts created by students in the seminar. (A) is one of a series of posts from an Instagram campaign to educate teachers about a new method for assessing the \textit{object-orientedness} of computer programs. (B) is an excerpt from an infographic poster on tips and pitfalls in various steps of the hiring pipeline in the technology sector.}
    \label{fig:impressions}
\end{figure*}

Fig.~\ref{fig:impressions} shows communication artifacts of two students (the texts are in German, the language in which the seminar took place, and it is not necessary to understand them to get an impression of the results).
First, we see (A) a post from a proposed Instagram campaign to inform teachers about a new method that could be used to assess the object-orientedness of a student-created computer program.
The student's communication strategy dealt with the paper by Tempero et al.~\cite{Tempero:2018:Objects} that discusses using the number of objects created in a program at runtime as a basis for evaluation of design decisions.
The Instagram post shown summarizes the motivation for the research paper.
Second, one can see (B) an excerpt of an infographic poster on tips and pitfalls in various steps of the hiring pipeline in the technology sector.
The communication strategy behind the poster deals with the paper by Behroozi et al.~\cite{Behroozi:2020:Hiring} on improving the technical interview process.
The target audience of the poster was identified as software companies facing problems in hiring applicants and seeking to improve their hiring process.
In both examples, we can see that in addition to text, a lot of work was done with colors and images to make the presentation more lively and appealing.
The seminar participants used public domain images that they found on the Internet.

All the students put a lot of thought into their ideas, and discussed them in terms of their potential effectiveness.
It was interesting to see which audiences the students identified for each research finding and how they tailored their communication strategy to each audience.
For example, one student looked at the paper by Minelli et al.~\cite{Minelli:2015:IKnowWhat} that investigated how developer work time is divided between activities.
She conceptualized and created a one-minute humorous video to communicate realistic expectations of the profession to prospective software engineering students.
The video could then be shared on the platforms that potential freshmen use.

\section{Discussion and Lessons Learned}

All seven students who completed the seminar participated in a brief evaluation of the course.
The evaluation took place via an anonymous online survey four days after the final presentations.
Participation in the survey was voluntary and included a consent form for dissemination of responses.
In the following, we combine student feedback from this evaluation with our own experiences into lessons learned that may help us and others in the future implementation of a similar course.

Five students identified as male, one as female, and one wished not to indicate their gender.
For four of the participants, the interest in the seminar topic was increased after the event compared to the interest before the semester started.
Two students maintained their high level of interest in science communication, and only one student's interest in the topic decreased slightly during the course.

\subsection{Assign papers early and give students the choice}

We asked students if they had any comments about the pre-selection or assignment process of the papers they were required to work on in the seminar.
One student felt it was very well done because \enquote{attention was paid to personal preference in paper selection, which made it easier to focus on science communication instead of having to struggle with a paper whose content one can't identify with}.
It was also noted that the process was clear and simple, and that it was helpful to make the assignment as early as possible so that one could familiarize oneself with the paper early on.

One student's feedback further confirmed our assumption that at the time of the preference submission, we could not yet assume that the students had clear understandings of the paper contents: \enquote{I think the selection was good. I skimmed the different abstracts and decided on 3 topics. In the end, I did not have an exact idea of my paper, but that is not necessarily a negative thing. I think with the choices given, everyone found a paper of their interest}.
The lightning talks, which take place shortly after the paper assignment, should therefore also serve the purpose of motivating students to engage more intensively with their paper at an early stage.

\subsection{Creative freedom and concrete requirements are equally desired}

We described to the students on one page at the very beginning what the portfolio of a communication strategy should be about.
In doing so, we briefly described the mandatory parts, made suggestions for elective parts, and stated that particular attention should be paid to discussing the expected benefits of individual components of the portfolio for science communication.

From student feedback, we understand that this had not yet been specific enough.
Smaller details, such as the expectation of the portfolio's size in terms of page count, would \enquote{take pressure off [students] to fill pages and instead focus on doing quality work}.
In addition, a student expressed the desire for more theoretical input, so that while one should continue to argue for one's communication strategy, one's argument could at best be supported with empirical evidence of the strategy's effectiveness.
We gratefully acknowledge both points.
We think a lot of uncertainty can be taken away from students by the instructor sketching out an exemplary portfolio and providing students with additional reading material.
It could even be a good variation for a future seminar that some of the papers the students can choose from deal with science communication and thereby provide the theoretical input themselves.
The knowledge gained could then be used to suggest improvements to existing software engineering research papers and their presentations.

At the same time, concrete formulations of the requirements should not mean that students are restricted in their creative choices.
The high level of creativity was mentioned positively several times by the students in the evaluation: \enquote{The creative way in which the subject was approached meant that you really questioned EVERYTHING you did and how you presented it, and I will keep that in mind in the future. There was a lot of freedom for your own ideas, appreciation and few constraints}.
Another student, when asked what they found particularly good about the seminar, described \enquote{that you learn that there is not one exact presentation style. And that everyone must adapt their slides to their own style of speaking}.
As a consequence of the few constraints, this student further highlights that \enquote{everyone had their freedom and each presentation was different, which was exactly the goal}.

\subsection{It only works if the seminar is a safe place for everyone}

We believe that good communication skills are important for every researcher and that appropriate training must be provided already for students.
Yet, there are cultural differences in the acceptance and handling of open criticism, for example when it is voiced by listeners of one's own presentation.
We must further not forget that computer science programs are currently predominantly attended by male students, and individuals of other genders face disproportionately higher hurdles to succeed in this environment.
Both of these aspects present the course instructor with what we believe to be the greatest challenge in the context of this seminar, namely, creating a safe and inclusive space in which all course participants receive and choose to receive fair and constructive feedback.
While a discrimination-free space should be the goal of any course, we consider that the course instructor of a seminar in which students move out of their comfort zone and, intentionally, make themselves the target of verbal criticism, has a special responsibility to be aware of these realities.

We can only encourage instructors to try and engage in dialogue with the seminar participants.
Overall, we had great pleasure in accompanying the students through this first seminar run.
Our impression was that the students were very open-minded and treated each other very respectfully, which led to a constructive feedback culture and apparent fun.
This was also reflected in the course evaluation, in which the seminar participants repeatedly praised the atmosphere and feedback from all other participants.

\subsection{Transferability to other disciplines and settings}

The learning objectives (\ref{sec:learning_objectives}) of our science communication seminar for computer science students should be easily transferable to other research disciplines.
The motivation for offering such a seminar should in most cases coincide with the one we have elaborated in the first pages of this paper.

Yet, it may be that the parameters for courses at other universities are different.
First, from a student perspective, our seminar stretched over a period of about 20 weeks (see Fig.~\ref{fig:overall_structure}).
Three to four weeks of this could be saved up to the time of the paper assignment, and two to three more in the period between lightning talks and portfolio submission.
This would still make the seminar as described take a little over three months.
For those instructors interested in ideas for a shorter alternative, we recommend, e.g., the work of Clark et al.~\cite{Clark:2016:ScienceEdu}.
In their paper, two concepts are being tested in which (PhD) students have to communicate their scientific work to middle school students.
This would not only improve the graduate students' science communication skills, but also provide science education for middle school students.

Second, one may not always have the opportunity to offer a brand-new seminar.
Luckily, Brownell et al.~\cite{Brownell:2013:SciComTeachStudents} describe an example of \enquote{infusing science communication training into the curriculum} and argue that \enquote{science communication skills need not be taught in stand-alone electives, but can be integrated effectively into lecture-based courses}.
Communication exercises could be incorporated into courses on basic science content.
For example, students could summarize the scientific literature covered in the course in the form of layperson-directed newspaper articles.
Brownell et al.~\cite{Brownell:2013:SciComTeachStudents} even went so far as to invite panels of science journalists and writers to critique students' work and thereby motivate them.
A few of the summaries were published in suitable forums.

So, for those who resonate with our motivation to improve science communication and would like to make a similar offer to their own students, but are limited in what they can do, there are alternative teaching options.
We deliberately chose to implement the seminar described because it offered students a variety of opportunities to try things out, including but not limited to writing a news article.

\section{Conclusion}

As software engineering researchers, we work on very different problems and with different intentions.
What we have in common is certainly the joy of finding answers to unsolved problems and enriching other people's lives with new insights.
The logical consequence is to grant access to these insights to the people we intend to help with our research.
Science communication offers the necessary tools for this purpose.
As with any tool, though, it is necessary to learn how to use it.

We experienced that university students can not only learn to reflect on the concrete way of communicating scientific findings, but also have a lot of fun doing so.
The first run of a science communication seminar in computer science courses at our university was already a success.
We were therefore keen to share our experience and encourage other instructors to run a seminar similar to ours.

To summarize, in this paper we have described the design of a science communication seminar in which students put themselves in the role of a researcher who, shortly after publication, has to draw attention to their paper and communicate its contents.
We turned the students' feedback into lessons learned and discussed them so that comparable seminars can be offered at other universities as well.
Collectively, university instructors can prepare the next generation of computer science researchers to not only be experts in their field, but to also effectively communicate research findings.
And whatever the future generation of computer science researchers may understand by the term: with good science communication, they are then very likely to get a bit closer to their goal of having \emph{impact}.

\bibliographystyle{IEEEtran}
\bibliography{bibliography}

\end{document}